\pdfoutput=1
\documentclass[10pt,twocolumn]{article}

\usepackage[letterpaper,margin=0.72in,columnsep=0.24in]{geometry}
\usepackage{microtype}
\usepackage{graphicx}
\usepackage{booktabs}
\usepackage{amsmath,amssymb}
\usepackage{xspace}
\usepackage{xcolor}
\usepackage{tikz}
\usetikzlibrary{arrows.meta,positioning,fit}
\usepackage{algorithm}
\usepackage{algorithmic}
\usepackage{hyperref}

\hypersetup{
  colorlinks=true,
  linkcolor=blue!45!black,
  citecolor=blue!45!black,
  urlcolor=blue!45!black,
  pdftitle={CacheRoute: Planned Prefix-Affinity Routing for Large-Scale LLM Serving},
  pdfauthor={Huang Cheng},
  pdfsubject={Large language model serving, prefix caching, and request routing},
  pdfkeywords={LLM serving, prefix caching, request routing, request scheduling}
}

\newcommand{\system}{\textsc{CacheRoute}\xspace}
\newcommand{\flatlb}{\textsc{Flat-LB}\xspace}
\newcommand{\qps}{\,QPS\xspace}

\title{\textbf{CacheRoute: Planned Prefix-Affinity Routing\\for Large-Scale LLM Serving}}
\author{Huang Cheng\\
Meta, Menlo Park, California, USA\\
\texttt{huangcheng@meta.com}}
\date{August 2026}

\begin{document}

\twocolumn[
\begin{@twocolumnfalse}
\maketitle
\begin{abstract}
Prefix caching avoids prefill only when a repeated request returns to a server that still holds the prefix KV.  Cache-blind balancing disperses that reuse; fixed affinity preserves it but can overload a server.  \system resolves this tradeoff with a periodic routing plan.  It admits high-rate keys to a stable warm set and places their assignments by expected load.  Hot keys may use more than one destination, although every key in our primary semi-synthetic aggregate uses exactly one.  On Llama-3.3-70B in fp8 across 60 H100 GPUs, \system sustains $176\!\pm\!11$\qps at a 3.5-s p99 SLO, $2.3\times$ the strongest of five baselines.  Served KV-cache hit rate rises from $64.1\!\pm\!1.3\%$ under cache-blind balancing to $93.2\!\pm\!0.5\%$.  A second semi-synthetic aggregate and controlled 8B and burst experiments separate the effects of affinity and placement.  Two 32B workloads provide the counterexamples: when affinity recovers too little KV work, its residual load skew reduces or erases the improvement.  We therefore recommend gating any deployment with a shadow replay rather than enabling affinity from workload statistics alone.
\end{abstract}
\vspace{0.15in}
\vspace{0.28in}
\end{@twocolumnfalse}
]

\section{Introduction}

Prefix caching is now standard in LLM serving engines~\cite{kwon2023vllm,zheng2024sglang}, but the cache cannot control where a request lands.  A cache-blind balancer spreads a recurring prefix across many destinations and lengthens the return time to each cache.  Pinning a key to one destination restores locality, but maps key skew directly onto server queues.  Cache-aware routers instead react to current cache and load state~\cite{srivatsa2024preble,yuan2026dualmap}; once a destination fills, however, spilling the next request also forfeits the warm prefix.

This problem arises in multi-tenant conversational assistants---for example, customer-support chatbots---where each business carries a stable, reusable context across a multi-turn conversation.  Such skewed, prefix-reusing serving is common across LLM serving systems and not specific to any one provider.  A stable per-business context recurs across conversations, while the default balancer often sends successive requests to different model servers.  Request rates also vary sharply across businesses, which rules out simple stickiness.  The router must keep reusable prefixes local without creating a server whose queue sets the fleet-wide p99.

\system builds a routing table from measured per-key rates.  It admits the highest-rate keys under a warm-prefix slot allocation, gives an overloaded key enough destinations to cap its expected load per destination, and places the resulting assignments with longest-processing-time-first (LPT) list scheduling.  Unadmitted traffic falls back to cache-blind power-of-two choices.  The table remains fixed for a control interval, favoring cache stability over per-request remapping.

Assignment and replication play different roles.  In the primary distribution, even the hottest key falls below the single-destination load target; every assignment count is one.  The 70B result therefore measures planned single-copy affinity, top-rate admission, and balanced placement.  Only the controlled 8B experiments inject keys hot enough to exercise replication.

The main experiment serves Llama-3.3-70B in fp8 on 30 tensor-parallel-2 destinations (60 H100 GPUs).  Across five paired runs, \system sustains $176\!\pm\!11$\qps at p99$\le3.5$\,s; the strongest baseline, Preble, sustains $76\!\pm\!11$\qps.  At 100 offered QPS, \system's p99 is 1.8\,s, compared with 3.8--8.5\,s for the five baselines.  On a second distribution, the advantage is $1.6\times$ at the wider active set.  The 8B ablation explains the gap: affinity raises KV hit rate, and LPT placement reduces imbalance from $3.46\times$ to $1.24\times$, moving the measured SLO knee from 240 to at least 500\qps.

We make three contributions:
\begin{itemize}
  \item a prefix-affinity router that plans cache locality and expected load together;
  \item a six-policy hardware evaluation centered on a 70B model and 60 H100 GPUs, with paired runs, a second workload distribution, and burst experiments; and
  \item a measured operating envelope that includes loss and tie regimes, plus evidence for using shadow replay instead of an analytic residency predictor.
\end{itemize}

Top-rate admission and LPT are established techniques; we do not claim a new scheduling primitive.  The contribution lies in the routing design, its large-scale measurements, and the conditions under which operators should leave the ordinary load balancer in place.

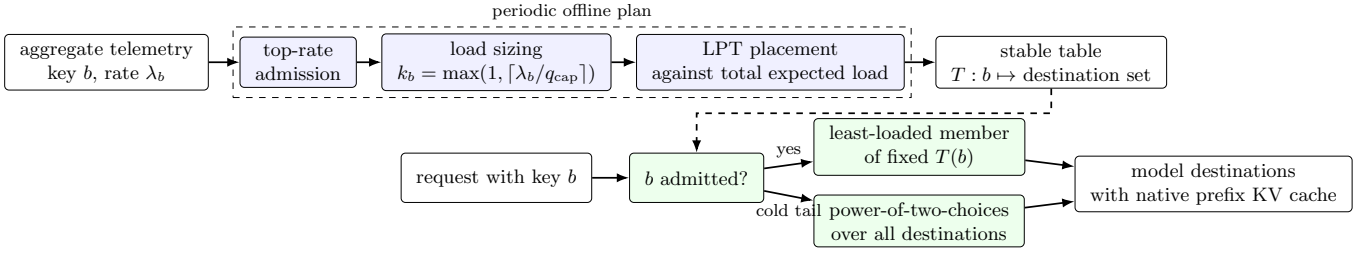
\begin{figure*}[t]
\centering
\resizebox{\textwidth}{!}{%
\begin{tikzpicture}[
  font=\small,
  box/.style={draw, rounded corners=2pt, align=center, minimum height=8mm, inner xsep=7pt},
  stage/.style={box, fill=blue!6},
  run/.style={box, fill=green!7},
  arr/.style={-{Latex[length=2mm]}, thick},
  note/.style={font=\footnotesize, align=center}
]
\node[box] (rates) {aggregate telemetry\\key $b$, rate $\lambda_b$};
\node[stage, right=5mm of rates] (admit) {top-rate\\admission};
\node[stage, right=4mm of admit] (size) {load sizing\\$k_b=\max(1,\lceil\lambda_b/q_{\rm cap}\rceil)$};
\node[stage, right=4mm of size] (place) {LPT placement\\against total expected load};
\node[box, right=5mm of place] (table) {stable table\\$T:b\mapsto$ destination set};
\draw[arr] (rates)--(admit);
\draw[arr] (admit)--(size);
\draw[arr] (size)--(place);
\draw[arr] (place)--(table);
\node[draw, dashed, fit=(admit)(size)(place), inner sep=3pt, label={[note]above:periodic offline plan}] {};

\node[box, below=10mm of size] (req) {request with key $b$};
\node[run, right=6mm of req] (selected) {$b$ admitted?};
\node[run, right=8mm of selected, yshift=5mm] (fixed) {least-loaded member\\of fixed $T(b)$};
\node[run, right=8mm of selected, yshift=-7mm] (fallback) {power-of-two-choices\\over all destinations};
\node[box, right=8mm of fixed, yshift=-6mm] (servers) {model destinations\\with native prefix KV cache};
\draw[arr] (req)--(selected);
\draw[arr] (selected)--node[note,above] {yes} (fixed);
\draw[arr] (selected)--node[note,below] {cold tail} (fallback);
\draw[arr] (fixed)--(servers);
\draw[arr] (fallback)--(servers);
\draw[arr, dashed] (table.south) -- ++(0,-4mm) -| (selected.north);
\end{tikzpicture}
}
\caption{\system separates a periodic, rate-aware assignment from fast request dispatch.  The main distribution has $k_b=1$ for every key, so its 70B improvement comes from admission, stable single-copy affinity, and balanced placement; multiple destinations are exercised only in the synthetic-whale mechanism study.  Cache residency is measured, not guaranteed by the plan.}
\label{fig:overview}
\end{figure*}

\section{Workload and Failure Mode}
\label{sec:workload}

\subsection{Semi-synthetic aggregate workload}

The workload models a multi-tenant conversational-assistant service (e.g., customer-support chatbots), where each business carries a stable, reusable context.  Our evaluation uses a \emph{semi-synthetic} workload: a synthetic request stream designed to mimic aggregate workload characteristics---per-key rates, inter-arrival gaps, and popularity distribution---derived from de-identified operational serving telemetry.  No raw requests, conversation content, user data, or business identifiers are used.  The primary summary contains 128,824 opaque business keys and has a Gini coefficient of 0.756.  Roughly 4\% of keys account for 47\% of requests, yet no key contributes more than 0.3\%.  The median per-key inter-arrival coefficient of variation is 1.93, and 80.8\% of requests belong to multi-turn threads.

Prompts average about 1.2K input tokens, 90\% of which precede the current turn.  Most of those tokens are not a routing opportunity.  The global template is common to all businesses and warms on every destination; a retrieved few-shot block changes often enough to break the exact prefix in 67\% of consecutive request pairs.  The reusable, key-specific segment is the per-business context, about 180 tokens or 15\% of the prompt.  We route on the business key and benefit only when that segment returns before eviction.

We evaluate two independently derived semi-synthetic aggregates and several explicitly labeled synthetic workloads.  The aggregates retain only replay statistics, including opaque-key rates and inter-arrival gaps.  Neither the paper nor its artifacts contain request text, user content, business names, or business identifiers.  We reserve synthetic traffic for mechanism and boundary tests, where controlling skew, burstiness, or prefix reuse matters more than matching a full workload.

\subsection{Why balancing and stickiness both fail}

For a business with rate $\lambda_b$ sprayed uniformly across $R$ destinations, the mean time between visits to one destination scales as $R/\lambda_b$.  Increasing the fleet can thus make a prefix \emph{colder} even though total cache capacity grows.  If this revisit time exceeds the effective eviction time, every request repeats prefill work.

Pure affinity moves to the other extreme.  A consistent hash keeps a prefix local, but the busiest hash bucket inherits the workload skew.  Tail latency then follows the hottest destination rather than the fleet average.  Bounded hashing and reactive reuse-versus-load policies reduce this problem, but may spill a request to a cold destination after a cap binds.  The practical need is a stable many-key assignment computed with the rate distribution in view.

Deployment turns on a measurable tradeoff: whether the \emph{recoverable} hit-rate increase outweighs the residual load skew.  Model size, precision, prefix length, offered load, active-set breadth, and cache size all shift that balance.  As Section~\ref{sec:envelope} shows, a higher cache-hit rate can still accompany lower capacity.

\section{CacheRoute}
\label{sec:design}

\subsection{Planning objective}

Let businesses $b\in B$ have estimated rates $\lambda_b$, and let the serving deployment have $R$ destinations.  A destination is one independently routable model-server group; in the 70B deployment it is a tensor-parallel-2 pair.  The planner produces a table $T(b)$ for a fixed control interval.

\paragraph{Load-based assignment count.}
We calibrate $q_{\mathrm{cap}}$ from a single destination's latency/load knee.  A selected business receives
\begin{equation}
  k_b=\max\left(1,\left\lceil\lambda_b/q_{\mathrm{cap}}\right\rceil\right)
  \label{eq:kb}
\end{equation}
destinations, making its expected load per assignment at most $q_{\mathrm{cap}}$.  This is a load-control rule, not an eviction threshold or a cache-residency guarantee.

\paragraph{Warm-set admission.}
The stable per-business prefixes are similar in size, so the planner represents cache allocation as $C=R W$ prefix slots, with $W$ slots per destination.  It considers keys in decreasing $\lambda_b$ and admits them while $\sum k_b\le C$.  This equal-slot model does not cover workloads with heterogeneous reusable-prefix sizes; those require byte-aware measurements and admission.

\paragraph{Placement and dispatch.}
Each admitted key contributes $k_b$ jobs of size $\lambda_b/k_b$.  After initializing destinations with their expected cold-tail share, the planner visits admitted keys in decreasing-rate order and assigns each job to the least-loaded eligible destination.  The procedure follows classic LPT scheduling~\cite{graham1969bounds}, but the standard approximation bound does not apply to our split jobs and distinct-destination constraint.  During the control interval, admitted traffic chooses the least-loaded member of its fixed $T(b)$; other traffic uses power-of-two choices across the fleet~\cite{mitzenmacher2001power}.

\begin{algorithm}[t]
\caption{Periodic routing-table construction}
\label{alg:plan}
\begin{algorithmic}[1]
\REQUIRE rates $\{\lambda_b\}$, $R$, $q_{\mathrm{cap}}$, slot allocation $C$
\FOR{each key $b$}
  \STATE $k_b\leftarrow\max(1,\lceil\lambda_b/q_{\mathrm{cap}}\rceil)$
\ENDFOR
\STATE sort keys by decreasing $\lambda_b$; admit while $\sum k_b\le C$
\STATE initialize $L[r]$ with expected cold-tail load, $r\in[1,R]$
\FOR{each admitted $b$ in decreasing $\lambda_b$}
  \STATE $T(b)\leftarrow k_b$ distinct destinations with smallest $L$
  \FOR{each $r\in T(b)$}
    \STATE $L[r]\leftarrow L[r]+\lambda_b/k_b$
  \ENDFOR
\ENDFOR
\STATE dispatch admitted $b$ within $T(b)$; otherwise use flat-LB
\end{algorithmic}
\end{algorithm}

The planner runs in $O(|B|\log |B|+\sum_b k_b\log R)$ and runs off the request path.  Measured construction time is 345\,ms at $R=30$ for 128,824 businesses.  The serving path is a table lookup followed by a load comparison within a small assignment set.

\subsection{Operational semantics}

The table stays fixed within a control interval.  Stability avoids per-request cap thrashing, but drift can make the assignment stale.  Installing a replacement table also starts some assignments cold.  Section~\ref{sec:operations} measures both effects.

Admitted and cold-tail prefixes share the engine's native cache and eviction policy.  \system neither reserves nor migrates KV blocks and offers no analytic residency guarantee.  Before enabling a table, we shadow-replay it and measure served token-weighted KV hit and latency.

For the primary distribution, $\max_b\lambda_b<q_{\mathrm{cap}}$ and therefore $k_b=1$ for every business.  Algorithm~\ref{alg:plan} reduces to top-rate admission followed by balanced, single-copy placement.  Replication remains available for overload protection but does not contribute to the main 70B result.

\section{Experimental Method}
\label{sec:method}

\paragraph{Implementation.}
Our client-side routing harness bypasses the serving deployment's aggregate load balancer and sends each request according to one policy.  An experiment identifier in request metadata attributes served, token-weighted KV-cache hits to the correct run.  Policies use the same serving stack, request stream, warm-up exclusion, offered-load ladder, and health criteria.  The routing layer neither changes model execution nor reserves cache capacity.

\paragraph{Testbeds.}
The flagship system is Llama-3.3-70B in fp8, served by 30 tensor-parallel-2 destinations on 60 H100 GPUs.  Each destination exposes 40,071 measured KV blocks (about 641K tokens).  The mechanism testbed is Llama-3.1-8B-Instruct in bf16 on 30 single-H100 destinations.  We use a 70B fp16 run only as supporting cache-pressure evidence, not for the capacity headline.

\paragraph{Workloads.}
The main replay draws opaque keys from a semi-synthetic peak-hour aggregate distribution.  A second, independently collected semi-synthetic aggregate distribution tests distribution shift.  Poisson, CV-matched Gamma, and moving-block-bootstrap arrival processes test burst sensitivity.  The 8B study uses a semi-synthetic aggregate distribution except where we explicitly inject synthetic whales.

\paragraph{Policies.}
We compare against \flatlb{} (power-of-two-choices), sticky consistent hashing~\cite{karger1997consistent}, consistent hashing with bounded loads (CHWBL)~\cite{mirrokni2018chwbl}, a DualMap-style two-candidate cache/load policy~\cite{yuan2026dualmap}, and a Preble-style prefix-history and live-load policy~\cite{srivatsa2024preble}.  The latter three are reimplemented in our harness under a common interface; our absolute results should not be read as reproductions of their original systems.  CHWBL uses $\epsilon=0.25$, selected by an offline sweep, and the Preble-style policy uses a 1.5 load-cap factor.

\paragraph{Metrics and statistics.}
SLO capacity is the highest offered-QPS ladder point with p99 latency at or below the stated threshold and failure rate at most 5\%.  We compute the threshold independently per run and report its mean; a value at the ladder maximum is right-censored.  The flagship top-$K$128 result uses five paired seeds $\{2,3,4,5,6\}$ and reports mean $\pm$ 95\% Student-$t$ confidence intervals.  The top-$K$256 confirmation uses eight seeds.  The secondary-distribution and 8B studies use three paired seeds unless stated otherwise; their wider intervals warrant interpreting small differences as noise.  KV-hit is served and token-weighted, not a simulator prediction.

\section{Evaluation}
\label{sec:evaluation}

We begin with the large-scale capacity result, then isolate the mechanism, examine workloads where affinity loses or ties, and derive the measurements required before enabling a routing plan.

\subsection{70B on 60 H100 GPUs}
\label{sec:flagship}

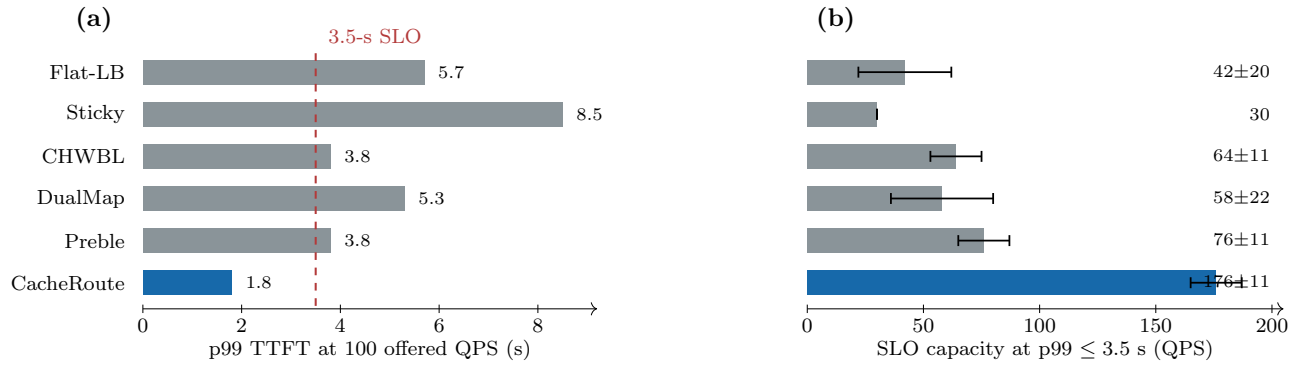
\begin{figure*}[t]
\centering
\resizebox{0.98\textwidth}{!}{
\definecolor{crblue}{HTML}{1769AA}
\definecolor{basegray}{HTML}{8A9499}
\definecolor{slored}{HTML}{B33A3A}
\begin{tikzpicture}[font=\footnotesize]
  \begin{scope}[xscale=0.68,yscale=0.58]
    \node[anchor=west,font=\bfseries] at (-1.55,6.25) {(a)};
    \foreach \name/\yy in {Flat-LB/5,Sticky/4,CHWBL/3,DualMap/2,Preble/1,CacheRoute/0}
      \node[anchor=east] at (-0.18,\yy) {\name};

    \foreach \value/\yy in {5.7/5,8.5/4,3.8/3,5.3/2,3.8/1}
      \fill[basegray] (0,\yy-0.29) rectangle (\value,\yy+0.29);
    \fill[crblue] (0,-0.29) rectangle (1.8,0.29);
    \foreach \value/\yy in {5.7/5,8.5/4,3.8/3,5.3/2,3.8/1,1.8/0}
      \node[anchor=west,font=\scriptsize] at (\value+0.10,\yy) {\value};

    \draw[slored,dashed,line width=0.8pt] (3.5,-0.55)--(3.5,5.55);
    \node[slored,anchor=south west] at (3.58,5.48) {3.5-s SLO};
    \draw[->] (0,-0.62)--(9.20,-0.62);
    \foreach \xx in {0,2,4,6,8}
      \draw (\xx,-0.72)--(\xx,-0.52) node[below=2pt] {\xx};
    \node at (4.6,-1.60) {p99 TTFT at 100 offered QPS (s)};
  \end{scope}

  \begin{scope}[xshift=9.15cm,xscale=0.032,yscale=0.58]
    \node[anchor=west,font=\bfseries] at (0,6.25) {(b)};

    \foreach \value/\yy in {42/5,30/4,64/3,58/2,76/1}
      \fill[basegray] (0,\yy-0.29) rectangle (\value,\yy+0.29);
    \fill[crblue] (0,-0.29) rectangle (176,0.29);

    \foreach \lo/\hi/\yy in {22/62/5,30/30/4,53/75/3,36/80/2,65/87/1,165/187/0} {
      \draw[line width=0.7pt] (\lo,\yy)--(\hi,\yy);
      \draw[line width=0.7pt] (\lo,\yy-0.13)--(\lo,\yy+0.13);
      \draw[line width=0.7pt] (\hi,\yy-0.13)--(\hi,\yy+0.13);
    }
    \foreach \label/\yy in {42$\pm$20/5,30/4,64$\pm$11/3,58$\pm$22/2,76$\pm$11/1,176$\pm$11/0}
      \node[anchor=east,font=\scriptsize] at (203,\yy) {\label};

    \draw[->] (0,-0.62)--(205,-0.62);
    \foreach \xx in {0,50,100,150,200}
      \draw (\xx,-0.72)--(\xx,-0.52) node[below=2pt] {\xx};
    \node at (102.5,-1.60) {SLO capacity at p99 $\leq3.5$ s (QPS)};
  \end{scope}
\end{tikzpicture}}
\caption{Primary 70B fp8 result on 30 TP2 destinations (60 H100), top-$K$128, five paired seeds.  \textbf{(a)} Measured p99 TTFT at the common 100-QPS operating point; only \system is below the primary 3.5-s SLO.  \textbf{(b)} Per-seed SLO-capacity knees (mean and reported 95\% Student-$t$ CI); \system reaches $176\!\pm\!11$\qps, $2.3\times$ the strongest baseline.}
\label{fig:70b-main}
\end{figure*}

Table~\ref{tab:70b} is the primary result; Figure~\ref{fig:70b-main} visualizes the common-load p99 and the primary SLO-capacity knee.  At p99$\le3.5$\,s, \system sustains $176\!\pm\!11$\qps, $2.3\times$ Preble's $76\!\pm\!11$ and $4.2\times$ \flatlb's $42\!\pm\!20$.  At a looser 5-s SLO, the gap narrows to $1.3\times$ over Preble (180 versus 140\qps).  At a tight 2-s SLO, \system is the only policy to pass any tested load and sustains 120\qps; all five baselines are left-censored below the 30-QPS minimum.

\begin{table}[t]
\caption{Llama-3.3-70B fp8, 30 TP2 destinations (60 H100), primary distribution, top-$K$128, five paired seeds.  Capacity is QPS at the SLO knee; KV-hit is served and token-weighted.}
\label{tab:70b}
\centering
\scriptsize
\setlength{\tabcolsep}{2.2pt}
\begin{tabular}{lrrrr}
\toprule
Policy & KV hit & Cap.@3.5s & Cap.@5s & p99@100 \\
\midrule
\flatlb       & $64.1{\pm}1.3$\% & $42{\pm}20$ & 80 & 5.7s \\
Sticky      & $87.3{\pm}2.4$\% & 30 & $42{\pm}20$ & 8.5s \\
CHWBL       & $75.6{\pm}0.8$\% & $64{\pm}11$ & $128{\pm}14$ & 3.8s \\
DualMap     & $88.7{\pm}1.9$\% & $58{\pm}22$ & $84{\pm}32$ & 5.3s \\
Preble      & $72.0{\pm}0.7$\% & $76{\pm}11$ & 140 & 3.8s \\
\system & $\mathbf{93.2{\pm}0.5}$\% & $\mathbf{176{\pm}11}$ & \textbf{180} & \textbf{1.8s} \\
\bottomrule
\end{tabular}
\end{table}

Cache and queue measurements explain the capacity gap.  \system records a 93.2\% served KV-hit rate, 29.1 points above \flatlb.  Sticky and DualMap also recover substantial reuse, but their p99s at 100 offered QPS are 8.5 and 5.3\,s; locality alone does not help once a queue dominates the tail.  Preble and CHWBL keep the queues more even, but their cache-hit rates fall to 72.0\% and 75.6\%.  \system occupies the useful middle: high reuse without a comparable hot spot.

With a top-$K$256 active set, a dedicated eight-seed run places the 3.5-s knees at 80\qps for \system and 60\qps for DualMap ($1.33\times$); at 5\,s they are 120 and 100\qps.  Their KV-hit rates are 82\% and 60\%, respectively.  Failure rates at passing points remain between 0.8\% and 1.4\%.  An earlier high-failure observation did not reproduce in this confirmation run.

\paragraph{Second semi-synthetic distribution.}
Table~\ref{tab:second} repeats the experiment with an independently collected aggregate key-rate distribution on the same 70B fp8 fleet.  When top-$K$128 fits within the warm allocation, the three balanced cache-aware policies tie at 100\qps.  Expanding to top-$K$256 separates them: \system sustains 160\qps, compared with 100\qps for the best baseline.  The comparison is useful precisely because it does not reproduce every cell of Table~\ref{tab:70b}; the advantage appears only after the active set outgrows the warm allocation.

\begin{table}[t]
\caption{SLO capacity (p99$\le3.5$\,s) on the second distribution, 70B fp8/60 H100, three paired seeds.}
\label{tab:second}
\centering
\small
\begin{tabular}{lrr}
\toprule
Policy & top-$K$128 & top-$K$256 \\
\midrule
\flatlb & 30 & 30 \\
Sticky & 30 & 60 \\
CHWBL & \textbf{100} & 60 \\
DualMap & \textbf{100} & 100 \\
Preble & 60 & 60 \\
\system & \textbf{100} & \textbf{160} \\
\bottomrule
\end{tabular}
\end{table}

\subsection{What creates the improvement?}
\label{sec:mechanism}

The 8B testbed reaches saturation with less load and separates locality from balance.  Across the tested active-set sizes, \system never falls below \flatlb's measured SLO capacity (Table~\ref{tab:8b}).  Several results at top-$K\in\{16,32,64\}$ hit the 500-QPS sweep ceiling; those ties are right-censored, not evidence of equal capacity.  The largest resolved six-policy advantage is $1.39\times$ at top-$K$128 and 256.  At top-$K$2000, the warm allocation covers a smaller traffic share and \system's measured knee falls to 360\qps, still tied for the best result.

\begin{table}[t]
\caption{8B SLO capacity (p99$\le3.5$\,s), 30 H100, three paired seeds.  ``500'' is right-censored.}
\label{tab:8b}
\centering
\scriptsize
\setlength{\tabcolsep}{2.7pt}
\begin{tabular}{lrrrrrr}
\toprule
top-$K$ & \flatlb & Sticky & CHWBL & DualMap & Preble & \system \\
\midrule
16   & 500 & 100 & 500 & 100 & 500 & \textbf{500} \\
32   & 500 & 100 & 360 & 100 & 360 & \textbf{500} \\
64   & 500 & 160 & 360 & 160 & 360 & \textbf{500} \\
128  & 240 & 240 & 360 & 360 & 360 & \textbf{500} \\
256  & 160 & 240 & 240 & 360 & 360 & \textbf{500} \\
2000 & 100 & \textbf{360} & 160 & \textbf{360} & 160 & \textbf{360} \\
\bottomrule
\end{tabular}
\end{table}

Table~\ref{tab:ablation} adds the components one at a time under a controlled synthetic-whale workload.  Affinity raises KV hit from 56\% to 88\%, but also raises imbalance to $3.46\times$; capacity stays at 240\qps.  Load-proportional replication reduces imbalance to $2.60\times$, yet random placement again leaves the knee unchanged.  Only after LPT placement brings imbalance down to $1.24\times$ does capacity reach the 500-QPS ceiling.  Affinity recovers the cached work, while placement makes that recovery usable at saturation.  The injected whales are the only reason replication appears in this ablation, so the row does not explain the primary result.

\begin{table}[t]
\caption{8B component ablation, synthetic whales, top-$K$128, three seeds.  Capacity 500 is right-censored.}
\label{tab:ablation}
\centering
\small
\begin{tabular}{lrrr}
\toprule
Configuration & KV hit & Imbalance & Capacity \\
\midrule
\flatlb & $56{\pm}1.9$\% & $1.00\times$ & 240 \\
$+$ affinity & $88{\pm}1.5$\% & $3.46\times$ & 240 \\
$+$ replication & $88{\pm}1.6$\% & $2.60\times$ & 240 \\
$+$ LPT (full) & $90{\pm}1.0$\% & $1.24\times$ & \textbf{500} \\
\bottomrule
\end{tabular}
\end{table}

\subsection{Operating envelope and negative results}
\label{sec:envelope}

Table~\ref{tab:negative} shows why affinity cannot be enabled unconditionally.  The 8B synthetic-whale workload offers many recoverable misses and favors \system.  For 32B aggregate workload~A, affinity moves KV hit only from 1.1\% to 11.8\%.  That improvement does not outweigh the remaining skew, and capacity falls to 0.50--0.67$\times$ \flatlb.  Aggregate workload~B reaches 8.5\% affinity hit and ties \flatlb at the 5-s SLO.  The two 32B workloads are de-identified semi-synthetic aggregates run with a different model configuration from the positive 70B case.  Model size by itself does not distinguish these outcomes.

\begin{table}[t]
\caption{Measured operating regimes.  Capacity multiplier is \system/\flatlb; the 8B range is a single-window synthetic mechanism result, while both 32B rows use semi-synthetic aggregate distributions.}
\label{tab:negative}
\centering
\small
\setlength{\tabcolsep}{3.0pt}
\begin{tabular}{lrrrl}
\toprule
Regime & Flat hit & Aff. hit & Cap. mult. & Result \\
\midrule
8B synthetic whales & 9.3\% & 77.0\% & 2--6$\times$ & win \\
Aggregate A--32B & 1.1\% & 11.8\% & 0.50--0.67$\times$ & lose \\
Aggregate B--32B & 0.8\% & 8.5\% & 1.0$\times$ & tie \\
\bottomrule
\end{tabular}
\end{table}

We use a short shadow replay as the deployment gate.  It preserves the candidate assignment without returning its responses to users.  The replay compares served KV hit, per-destination load, and p99 against \flatlb at one load below and one load near the current knee.  A cache-hit increase is not enough; the plan is enabled only when p99 or capacity improves.  We repeat the gate after changes to model, precision, context length, batching, cache allocation, or the active-set distribution.

\subsection{Burstiness and replanning}
\label{sec:operations}

\paragraph{Arrival burstiness.}
On 70B fp8/top-$K$128, a Gamma arrival process matched to the workload's marginal CV=1.9 reduces \system's 3.5-s capacity by one ladder step, from 180 to 160\qps, while KV-hit moves from 94\% to 90\%.  \flatlb remains at 30\qps.  In a separate, single-sweep comparison, a moving-block bootstrap of 22,639 measured inter-arrival gaps from semi-synthetic aggregate workload~A (block length 50, empirical CV=2.73) leaves \system at 180\qps and 93\% hit, while \flatlb becomes left-censored below 30\qps.  These are two-policy, one-active-set sensitivities, not claims about all baselines.  They show that the main improvement does not require smoothed arrivals.

\paragraph{Rate drift and remapping.}
We perturb the primary rate vector over six intervals with a lognormal random walk ($\sigma=0.3$) and compare fresh and one-interval-stale plans at 160\qps.  Recomputing LPT from scratch changes 94.5\% of key-to-destination sets, although only 1.1\% of keys change assignment count.  The stale plan loses 1.0 percentage point of KV hit and 192\,ms of p99 on average; the worst transition loses 3.0 points and 858\,ms.  Installing the fresh plan, meanwhile, produces a transient 13.6-point KV-hit drop during rewarming.  A deployment should therefore preserve placements until the measured staleness penalty exceeds the warm-up penalty.  We have not implemented a churn-aware replanner.

\paragraph{Why we do not predict residency analytically.}
We also tested a single-characteristic-time occupancy model against the 70B fp8 engine.  Component-level instrumentation passed seven isolation checks, and each TP2 destination exposed 40,071 measured KV blocks.  Prediction still missed served hit rate by 14.3 percentage points at the median and 44.7 points at p90.  The observed curve drops sharply and then plateaus, a shape outside the tested model family.  We neither size deployments with this model nor assign the shape to an engine mechanism without direct evidence.  This failure is why shadow replay remains part of the deployment procedure.

\section{Discussion and Limitations}

\paragraph{What generalizes.}
The planner requires a stable routing key, a reusable key-specific prefix, and rate estimates that remain useful for at least one cache-warming interval.  Another routing key---a tenant, a document collection, an agent, or an application---could replace the business key used here.  Because the router changes placement rather than model execution, it does not depend on a particular model architecture or cache implementation.

\paragraph{What does not yet generalize.}
Equal-slot admission assumes roughly uniform stable-prefix sizes.  We did not measure a byte-aware value function for heterogeneous long contexts and make no byte-knapsack or optimal-admission claim.  Cold-tail traffic shares the native cache and can evict admitted prefixes.  The aggregate replays preserve key popularity and, for the burst experiment, short-range gap correlation; they do not preserve exact per-key timestamps.  Finally, two distributions cannot represent every market or application.

Preble, DualMap, and CHWBL are common-harness reimplementations, not the original codebases.  They compare routing behavior under matched hardware but do not reproduce the published systems end to end.  Several 8B knees are right-censored at 500 QPS, so we report ratios only when the sweep resolves them.  The three-seed studies also have wide $t$ intervals; the five-seed flagship and eight-seed confirmation provide the stronger evidence.

\paragraph{Deployment checklist.}
An operator adopting \system would begin with aggregate key rates and a single-destination measurement of $q_{\mathrm{cap}}$.  After choosing a conservative warm-slot allocation, they would shadow-replay the plan and \flatlb{} at matched loads, comparing served KV hit, imbalance, p99, and failures.  Canary traffic would follow only if the plan wins, and the router would fall back to \flatlb{} when the active set or measured p99 leaves that envelope.  None of these steps requires request content.

\section{Related Work}

PagedAttention/vLLM and SGLang make prefix KV reuse efficient within one serving engine~\cite{kwon2023vllm,zheng2024sglang}.  \system instead decides which destination receives the request.

Preble scores prefix reuse against live load, while DualMap combines static and dynamic mappings~\cite{srivatsa2024preble,yuan2026dualmap}.  Consistent hashing and CHWBL provide stable affinity with explicit load control~\cite{karger1997consistent,mirrokni2018chwbl}.  \system constructs one global, rate-based table for a control interval.  Admission and LPT are established heuristics; the new evidence concerns their use in a prefix-affinity plan and the workloads on which that plan helps.

GeoCover studies a different form of resource-constrained placement~\cite{cheng2015geocover}: it selects roadside units for spatial network coverage, whereas \system assigns recurring prefixes to model destinations under serving load.  The algorithms and guarantees do not transfer between the two problems.  Their common ground is the allocation of limited infrastructure under skewed demand.

Mooncake and MemServe move or pool KV state across servers rather than preserve it through ingress affinity~\cite{hu2024memserve,qin2025mooncake}.  Llumnix migrates requests to repair imbalance~\cite{sun2024llumnix}; DistServe and Splitwise separate prefill and decode placement~\cite{zhong2024distserve,patel2024splitwise}.  These systems can still use an ingress plan to avoid prefill that routing would otherwise repeat.

\section{Conclusion}

\system plans prefix affinity and expected load once per control interval.  On a 70B fp8 model across 60 H100 GPUs, it reaches 93.2\% served KV hit and $2.3\times$ the capacity of the strongest baseline at a 3.5-s p99 SLO.  The negative workloads set an equally important boundary: affinity can reduce capacity when the recoverable prefix work is small.  Any real deployment would therefore measure its own boundary with shadow replay.

\clearpage
\appendix
\section*{Supplementary Material}
The supplement records experiment provenance, secondary runs, and negative results.  It adds no headline claims.  Each result is labeled as multi-run hardware, single-window hardware, or calibrated simulation.

\section{Workload Detail}

\subsection{Semi-synthetic aggregate statistics}

The routing key is an opaque business identifier, and the planner reads aggregate counts rather than request text.  Table~\ref{tab:workload} reports de-identified summaries derived from operational serving telemetry and distinguishes overall prompt structure from the segment whose residency the router can affect.

\begin{table}[H]
\caption{Primary workload characteristics.  ``Positional prefix'' is not the same as the business-specific routing lever.}
\label{tab:workload}
\centering
\scriptsize
\setlength{\tabcolsep}{4pt}
\begin{tabular}{lr}
\toprule
Property & Measured value \\
\midrule
Distinct business keys & 128,824 \\
Traffic Gini coefficient & 0.756 \\
Top-key share & $<0.3\%$ \\
Top 4\% share & $\approx47\%$ \\
Multi-turn request share & 80.8\% \\
Mean requests per thread & 2.64 \\
Median per-key inter-arrival CV & 1.93 \\
Mean input length & $\approx1.2$K tokens \\
Tokens preceding current turn & $\approx90\%$ \\
Stable per-business context & $\approx180$ tokens ($15\%$) \\
Pairs whose exact prefix breaks at RAG & 67\% \\
\bottomrule
\end{tabular}
\end{table}

Each prompt contains a global template, stable per-business context, retrieved few-shot block, and current turn.  The shared template warms regardless of routing.  Changes to the retrieved block often truncate exact reuse, leaving the per-business context as the stable routing-specific segment.  Another application would need to identify its own stable segment before using the same design.

\subsection{Data handling}

The workload inputs are aggregate per-key rates, prompt-length statistics, served cache counters, and inter-arrival gaps.  Raw requests, text, user content, names, and business identifiers are absent from both the paper and its artifacts.  The planner needs only $\{(b,\lambda_b)\}$ and live destination load; $b$ may be salted or ephemeral.  Synthetic workloads are labeled separately, and infrastructure measurements are reported only at experiment level, without host identifiers or per-request records.

\section{Experiment Provenance and Statistics}

\subsection{Study matrix}

Table~\ref{tab:provenance} lists the replication unit for each study.  Every seed defines a separate offered-load realization and warm-up exclusion.  Policies within a row share seeds and request distributions, so their capacity comparisons are paired.

\begin{table*}[t]
\caption{Experiment provenance.  ``HW'' denotes real H100 execution; ``sim'' denotes the calibrated offline simulator.  The synthetic-whale and calibrated-simulation results are supporting mechanism evidence, not broad generalization evidence.}
\label{tab:provenance}
\centering
\scriptsize
\setlength{\tabcolsep}{2.2pt}
\begin{tabular}{llllll}
\toprule
Study & Model/precision & Destinations & Workload & Repetitions & Evidence class \\
\midrule
Primary six-policy & Llama-3.3-70B/fp8 & 30 TP2 (60 H100) & primary aggregate, top-$K$128 & seeds 2--6 ($n=5$) & HW/core \\
Wide active set & Llama-3.3-70B/fp8 & 30 TP2 (60 H100) & primary aggregate, top-$K$256 & seeds 2--9 ($n=8$) & HW/core \\
Second distribution & Llama-3.3-70B/fp8 & 30 TP2 (60 H100) & independent aggregate, $K\in\{128,256\}$ & seeds 2--4 & HW/core \\
Precision pressure & Llama-3.3-70B/fp16 & 30 TP2 (60 H100) & primary aggregate & seeds 2--4 & HW/supporting \\
8B six-policy/ablation & Llama-3.1-8B/bf16 & 30 MP1 (30 H100) & aggregate or injected whales & seeds 2--4 & HW/mechanism \\
Burst--Gamma & Llama-3.3-70B/fp8 & 30 TP2 (60 H100) & marginal CV=1.9 & seeds 2--6 & HW/sensitivity \\
Burst--trace blocks & Llama-3.3-70B/fp8 & 30 TP2 (60 H100) & 22,639 gaps, block 50 & seeds 2--6 & HW/sensitivity \\
Drift/replan & Llama-3.3-70B/fp8 & 30 TP2 (60 H100) & six perturbed intervals & seeds 2--4 & HW/sensitivity \\
32B negative regimes & Qwen3-32B/bf16 & 30 MP2 (60 H100) & two semi-synthetic aggregates & matched runs & HW/negative \\
Outer-boundary sweep & 8B-calibrated & $R=30$ & synthetic prefix fraction/load & seeds 2--4 & sim/supporting \\
\bottomrule
\end{tabular}
\end{table*}

\subsection{Threshold construction}

For each run and policy, capacity is the largest offered-load point whose p99 meets the SLO and whose failure rate is at most 5\%.  Reported capacity is the mean of these per-run knees, rather than a confidence interval around one latency measurement.  A run that fails at the lowest load is left-censored; one that passes at the maximum is right-censored.  The primary 70B ladder is $\{30,60,80,100,120,140,160,180,220,300\}$ QPS, and the 8B ladder is $\{30,60,100,160,240,360,500\}$ QPS.

Continuous measurements use the sample mean and a two-sided 95\% Student-$t$ interval.  With $n=3$, $t_{0.975,2}=4.303$, so those intervals are necessarily wide; with $n=5$, $t_{0.975,4}=2.776$.  Experiments, not individual requests, are the replication units.

\section{Additional 70B Results}

\subsection{fp16 cache-pressure run}

Table~\ref{tab:fp16} checks locality under greater KV pressure.  No policy meets the fp8 study's 3.5-s SLO, so this run supplies no capacity multiplier and is not combined with the fp8 headline.

\begin{table}[H]
\caption{70B fp16, 30 TP2 destinations/60 H100, top-$K$128, three seeds.  Capacity is zero because all policies miss the fp8 study's 3.5-s SLO.}
\label{tab:fp16}
\centering
\scriptsize
\setlength{\tabcolsep}{2.5pt}
\begin{tabular}{lrr}
\toprule
Policy & Served KV hit & Capacity@3.5s \\
\midrule
\flatlb & $19.5\pm2.1\%$ & 0 \\
Sticky & $77.3\pm0.7\%$ & 0 \\
\system & $76.4\pm0.5\%$ & 0 \\
\bottomrule
\end{tabular}
\end{table}

At top-$K$256, KV-hit is approximately 10\% for \flatlb{} and 77\% for \system.  The comparison measures the effect of scattering at model scale; it does not show a cache-hit advantage over sticky routing.

\subsection{\texorpdfstring{Top-$K$256}{Top-K256} confirmation}

An earlier five-seed top-$K$256 sweep showed failures unrelated to load.  A dedicated eight-seed confirmation on the same fp8 fleet and a finer ladder did not reproduce them: failure rates for all \system seeds are 0.8--1.4\% at 30 and 60 QPS, comparable with both baselines.  Table~\ref{tab:k256} reports the confirmation.  We retain the earlier anomaly in the provenance but do not treat it as a repeatable policy reversal.

\begin{table}[H]
\caption{Dedicated 70B fp8/top-$K$256 confirmation ($n=8$).}
\label{tab:k256}
\centering
\small
\begin{tabular}{lrrr}
\toprule
Policy & KV hit & Cap.@3.5s & Cap.@5s \\
\midrule
DualMap & 60\% & 60 & 100 \\
\system & \textbf{82\%} & \textbf{80} & \textbf{120} \\
\bottomrule
\end{tabular}
\end{table}

\section{Arrival-Process Sensitivity}

The two arrival studies were run separately.  Table~\ref{tab:gamma} compares fixed-rate and marginal-CV-matched Gamma arrivals; Table~\ref{tab:blocks} adds a moving-block bootstrap in a new sweep.  Their Poisson knees differ (30 versus 60 QPS), so ratios are computed only within each table.

\begin{table}[H]
\caption{Gamma sensitivity, 70B fp8/top-$K$128, five seeds.  Capacity is at p99$\le3.5$\,s.}
\label{tab:gamma}
\centering
\small
\begin{tabular}{lrrrr}
\toprule
 & \multicolumn{2}{c}{\flatlb} & \multicolumn{2}{c}{\system} \\
\cmidrule(lr){2-3}\cmidrule(lr){4-5}
Arrival & Cap. & KV & Cap. & KV \\
\midrule
Poisson & 30 & 64\% & 180 & 94\% \\
Gamma, CV=1.9 & 30 & 63\% & 160 & 90\% \\
\bottomrule
\end{tabular}
\end{table}

\begin{table}[H]
\caption{Self-contained block-bootstrap sweep, 70B fp8/top-$K$128, five seeds.  The block trace has empirical CV=2.73.  Zero is left-censored below 30 QPS.}
\label{tab:blocks}
\centering
\small
\begin{tabular}{lrrrr}
\toprule
 & \multicolumn{2}{c}{\flatlb} & \multicolumn{2}{c}{\system} \\
\cmidrule(lr){2-3}\cmidrule(lr){4-5}
Arrival & Cap. & KV & Cap. & KV \\
\midrule
Poisson & 60 & 66\% & 180 & 93\% \\
Gamma, CV=1.9 & 60 & 66\% & 180 & 93\% \\
Trace blocks, CV=2.73 & 0 & 67\% & 180 & 93\% \\
\bottomrule
\end{tabular}
\end{table}

The bootstrap draws contiguous blocks of 50 from 22,639 measured inter-arrival gaps.  It retains short-range gap correlation, but neither the original global order nor each key's exact timestamp sequence.  We leave timestamp-exact replay to future work.

\section{Replanning Under Drift}

We evolve each key's rate for six intervals with a lognormal random walk ($\sigma=0.3$ per step) around the primary distribution.  Churn uses all five transitions; the 160-QPS hardware comparison uses three transitions with paired seeds.

\begin{table}[H]
\caption{Detailed replan measurements.  Cache-churn is $1-$Jaccard of consecutive assignment sets; count-churn is the fraction whose $k_b$ changes.  Stale losses compare the previous plan with a freshly computed plan.}
\label{tab:replan}
\centering
\scriptsize
\setlength{\tabcolsep}{1.8pt}
\begin{tabular}{lrrrrr}
\toprule
Transition & Cache churn & Count churn & KV loss & p99 loss & Warmup dip \\
 & & & (points) & (ms) & (points) \\
\midrule
$0\to1$ & 92.5\% & 1.6\% & 0.1 & $-119$ & 15.6 \\
$2\to3$ & 94.5\% & 0.0\% & 3.0 & $+858$ & 11.4 \\
$4\to5$ & 98.0\% & 1.6\% & $-0.1$ & $-162$ & 13.8 \\
\midrule
Mean & 94.5\% & 1.1\% & 1.0 & $+192$ & 13.6 \\
\bottomrule
\end{tabular}
\end{table}

Recomputing LPT from scratch accounts for the high cache churn: a key may keep the same assignment count but move to another equally loaded destination.  Placement hysteresis could avoid such moves, although we have not evaluated a churn-aware algorithm.  For now, installing a new table should depend on whether the measured stale-plan penalty exceeds the measured warm-up penalty.

\section{8B Sensitivity and Replication Scope}

\subsection{Load target and skew}

Table~\ref{tab:sensitivity} varies the load target and injected head share on the 8B testbed.  Under the base (non-whale) distribution, every key remains below even the smallest $q_{\mathrm{cap}}$, so $k_b=1$ throughout the sweep.  The unchanged result is a sanity check, not evidence that replication is insensitive to its load target.

\begin{table}[H]
\caption{8B sensitivity, $R=30$, top-$K$128, three seeds.  Capacities are ladder knees at p99$\le3.5$\,s; 500 is right-censored.}
\label{tab:sensitivity}
\centering
\scriptsize
\setlength{\tabcolsep}{1.8pt}
\begin{tabular}{llrrr}
\toprule
Axis & Setting & KV (CR/flat) & CR cap. & Mult. \\
\midrule
$q_{\mathrm{cap}}$ & 50 QPS  & $89\pm0.9\%/57\%$ & 360 & $1.50\times$ \\
$q_{\mathrm{cap}}$ & 100 QPS & $91\pm3.1\%/57\%$ & 360 & $1.50\times$ \\
$q_{\mathrm{cap}}$ & 200 QPS & $91\pm3.2\%/57\%$ & 360 & $1.50\times$ \\
$q_{\mathrm{cap}}$ & 500 QPS & $91\pm3.1\%/57\%$ & 360 & $1.50\times$ \\
\midrule
Injected skew & 5\% head  & $89\pm1.1\%/56\%$ & 500 & $3.12\times$ \\
Injected skew & 15\% head & $89\pm1.0\%/56\%$ & 500 & $2.08\times$ \\
Injected skew & 25\% head & $89\pm1.1\%/57\%$ & 500 & $2.08\times$ \\
Injected skew & 45\% head & $89\pm1.1\%/59\%$ & 500 & $2.08\times$ \\
\bottomrule
\end{tabular}
\end{table}

\subsection{Fleet size under fixed absolute load}

Fleet size produces a non-monotonic result in the synthetic-whale sweep (Table~\ref{tab:fleet}).  Because absolute QPS stays fixed, increasing $R$ lowers the repeat rate seen by each destination; at $R=60$, even targeted prefixes become cold.  A deployment whose load scales with fleet size may behave differently.

\begin{table}[H]
\caption{Synthetic-whale fleet-size sweep.  The $R=8$ ramp has no passing capacity point; $R=60$ has no winning scenario.}
\label{tab:fleet}
\centering
\scriptsize
\setlength{\tabcolsep}{2.3pt}
\begin{tabular}{lrrr}
\toprule
$R$ & Representative scenario & KV (CR/flat) & Result \\
\midrule
8 & 2 whales@3.0 & 45\%/12\% & tail/KV only \\
30 & 8 whales@3.0 & 70\%/14\% & $4.44\times$ \\
60 & 16 whales@3.0 & 32\%/24\% & no win \\
\bottomrule
\end{tabular}
\end{table}

\section{Negative Hardware Regimes}

The two negative regimes use Qwen3-32B in bf16 on 30 MP2 destinations (60 H100), de-identified semi-synthetic aggregates, and no injected whales.  They are separate workloads from the positive 70B fp8 case.  Table~\ref{tab:app-negative} reports measured knees rather than only ratios.

\begin{table}[H]
\caption{Negative and neutral hardware regimes.  A dash denotes an unreported threshold, not zero.}
\label{tab:app-negative}
\centering
\scriptsize
\setlength{\tabcolsep}{2.5pt}
\begin{tabular}{lrrrrr}
\toprule
 & \multicolumn{2}{c}{KV hit} & \multicolumn{2}{c}{Capacity@3.5s} & Cap.@5s \\
\cmidrule(lr){2-3}\cmidrule(lr){4-5}
Regime & flat & affinity & flat & CR & flat/CR \\
\midrule
Aggregate A--32B & 1.1\% & 11.8\% & 20 & 10 & 30/20 \\
Aggregate B--32B & 0.8\% & 8.5\% & -- & -- & 20/20 \\
\bottomrule
\end{tabular}
\end{table}

On aggregate A--32B, \system reaches $0.50\times$ \flatlb's capacity at 3.5\,s and $0.67\times$ at 5\,s.  The policies tie on aggregate B--32B at 5\,s.  Affinity adds only 8--11 percentage points of cache hit in these workloads, too little to offset modest skew.  This reversal motivates the shadow-replay gate used in the main paper.

\section{Analytic Residency Model: Negative Result}

Each TP2 destination in the 70B fp8 engine exposes 40,071 physical KV blocks, or approximately 641K tokens.  An earlier configuration note used 30,000 blocks; the ratio between the values is $1.34\times$.  Our results use the measured block count, not that ratio, as the relevant configuration.

Component instrumentation passes seven isolation tests, and the business-hit crossover remains stable across the eviction sweep (coefficient of variation 0.038 over 24 cells).  A single-characteristic-time occupancy model nevertheless misses served hit rate by 14.3 percentage points at the median and 44.7 points at p90; its rank agreement also falls below the deployment target.  The measured residency curve drops sharply and then plateaus, outside the tested model family.  We lack direct evidence to attribute that shape to batching, block quantization, or another engine mechanism.

Physical block capacity and aggregate rates are therefore insufficient for pre-deployment sizing on this engine.  The deployment interface includes a shadow replay that measures served KV hit and p99 for the candidate assignment.

\paragraph{Claim-to-evidence map.}
The main claims map to the following evidence:
\begin{itemize}
  \item The $2.3\times$ headline is supported by the 70B fp8, 60-H100, top-$K$128, five-seed hardware study; it is not a universal multiplier.
  \item The wider-active-set result is a separate eight-seed confirmation and is smaller ($1.33\times$ at 3.5\,s).
  \item The second distribution ties at top-$K$128 and reaches $1.6\times$ only at top-$K$256.
  \item The 8B ablation supports the affinity/placement decomposition; only synthetic whales exercise replication, while the primary workload has $k_b=1$.
  \item Burst robustness is a two-policy, one-active-set sensitivity and is not extrapolated to the other four baselines.
  \item The 32B loss/tie workloads establish a deployment gate, and the failed analytic predictor supports measuring residency rather than asserting a microarchitectural cause.
\end{itemize}

\end{document}